\documentclass[prb,showpacs,amsmath,amssymb,twocolumn]{revtex4}

\usepackage{graphicx}
\usepackage{epsfig}
\usepackage{dcolumn}
\usepackage{bm}

\begin{document}

\title{Valley Splitting Theory of SiGe/Si/SiGe Quantum Wells}

\author{Mark Friesen$^1$}
\email{Electronic mail:  friesen@cae.wisc.edu}
\author{Sucismita Chutia$^1$}
\author{Charles Tahan$^2$}
\author{S. N. Coppersmith$^1$}
\affiliation{$^1$Department of Physics, University of
Wisconsin-Madison, Wisconsin 53706, USA} 
\affiliation{$^2$Cavendish Laboratory, University of Cambridge, JJ Thomson Ave, 
Cambridge CB3 0HE, United Kingdom}

\begin{abstract}
We present an effective mass theory for SiGe/Si/SiGe
quantum wells, with an emphasis on calculating the valley splitting.  The theory
introduces a valley coupling parameter, $v_v$, which 
encapsulates the physics of the quantum well interface.
The new effective mass parameter is computed by means of a tight binding theory.
The resulting formalism provides rather simple
analytical results for several geometries of interest,
including a finite square well, a quantum well in an electric field, and
a modulation doped two-dimensional electron gas.  Of particular importance is
the problem of a quantum well in a magnetic field, grown on a miscut substrate.
The latter may pose a numerical challenge for
atomistic techniques like tight-binding, because of its two-dimensional nature.  
In the effective mass theory, however, the results are straightforward and analytical. 
We compare our effective mass results with
those of the tight binding theory, obtaining excellent agreement.
\end{abstract}


\maketitle

\section{Introduction}
Silicon heterostructures form the basis of numerous semiconductor technologies.  
To understand future silicon heterostructure devices that involve quantum effects, 
one must consider the quantum states associated with the degenerate valleys in the 
conduction band structure.  In a strained quantum well, the valley degeneracy is two-fold. 
This degeneracy is lifted by the singular nature of the quantum well interface, 
with characteristic energy splittings of order 0.1-1~meV for the case of
SiGe/Si/SiGe quantum wells.\cite{weitz96,koester97,khrapai03,lai04,pudalov,goswami06} 
Because this splitting is comparable in size to the Zeeman splitting, the valley states can
compete with spin states for prominence in quantum devices.\cite{eriksson04}
For emerging technologies like silicon spintronics 
\cite{rokhinson02,xiao04,zuticRMP,bradbury06,sellier06} and quantum 
computing,\cite{kane,vrijen,obrien01,friesen03} it is therefore crucial to
obtain a solid physical understanding of the valley physics, and to develop a 
predictive theory for the valley states.

The technological significance of valley states was recognized long before the 
current interest in quantum devices.\cite{kohn,AFS}  Early studies focused on  
bulk silicon, particularly on the electronic states of shallow donors. 
In this case, valley splitting was known to originate from the singular core of the donor
potential, known as the ``central cell."\cite{yubook}  In one theoretical approach,
the many-body interactions associated with
the central cell were projected onto a single-electron, effective mass (EM) framework,
allowing the energy spectrum of the low-lying hydrogenic states to be computed from 
first principles.\cite{pantelides,pantelides2}  Reasonable agreement with experiments was 
attained.  However, because of the complicated projection procedure,
a general purpose EM theory was not obtained until more 
recently.\cite{friesen05}

Related techniques were applied to the problem of valley splitting
in a semiconductor heterostructure.  There were several attempts to 
develop an EM theory, which is well suited for treating 
inhomogeneous conditions, including conduction band offsets 
and non-uniform electric fields associated with modulation 
doping and top-gates.  However, the resulting theories proved
controversial, and many important valley splitting problems remain unsolved.  

Atomistic approaches like tight-binding (TB) theory have recently emerged
as important tools for calculating the valley splitting. \cite{boykin04}  
These techniques have been applied to a range of heterostructure
geometries,\cite{boykin04,boykin04b,boykin05} providing crucial insights and
predictions for experiments that that may be difficult to implement.
For example, the valley splitting has been predicted to oscillate 
as a function of the quantum well width.\cite{boykin04}  These oscillations can be 
reduced, or even eliminated, by applying an electric field.  

Here, we develop an EM formalism, which corroborates the atomistic
results quite accurately, and which provides a simple
physical explanation for the intriguing oscillations.  Specifically, we show that the 
behavior occurs because of valley coupling interference between the top and bottom 
interfaces of the quantum well.  In an electric field, the wavefunction is
squeezed to one side of the quantum well, thereby eliminating the interference effect.
We also use the EM theory to move beyond simple one-dimensional (1D) geometries.  For
example, quantum wells grown on a miscut substrate represent an inherently 2D problem.  
Here, we show that interference effects also play a crucial role for such miscut geometries,
causing a strong suppression of the valley splitting at low magnetic fields.

The paper is organized as follows.  In Sec.~\ref{sec:approaches}, we review the two
predominant approaches to valley coupling.  In Sec.~\ref{sec:EMT}, we describe an 
extension to the conventional effective mass theory that provides a perturbative 
scheme to incorporate valley coupling.
In Sec.~\ref{sec:TB}, we use a tight binding theory
to calculate the new input parameter for the EM theory -- the valley coupling 
$v_v$ -- as a function of the conduction band offset $\Delta E_c$. 
In Sec.~\ref{sec:finite}, we apply the EM theory to a finite square well
geometry.  In Sec.~\ref{sec:Efield}, we obtain an analytical solution for a quantum well 
in an external electric field.  In Sec.~\ref{sec:2DEG}, we consider the experimentally 
important problem of a two-dimensional electron gas (2DEG).  
In Sec.~\ref{sec:steps}, we study the valley splitting in a magnetic field, 
when the quantum well is misaligned with respect to the crystallographic axes.
Finally, in Sec.~\ref{sec:conclusions}, we summarize our results and conclude.

\section{Effective Mass Approach}
\label{sec:approaches}
Earlier EM approaches for the valley splitting 
in a heterostructure include the ``electric breakthrough" theory of Ohkawa and 
Uemura,\cite{ohkawa77a,ohkawa77b,ohkawa78}
and the surface scattering theory of Sham and Nakayama.\cite{sham79}  A review of these
theories and other related work is given
in Ref.~[\onlinecite{AFS}].  The Ohkawa-Uemura formalism leads to a multi-valley  
EM theory based on a two-band model involving the lowest two conduction 
bands at the $\Gamma$ point of the Brillouin zone,
$|i\Gamma_1(S)\rangle$ and $| \Gamma_{15}(Z)\rangle$.\cite{ohkawa79}
The resulting bulk dispersion relation has a local maximum at the $\Gamma$ point,
and it exhibits two degenerate valleys, at roughly the correct positions in $k$-space.  
The sharp confinement potential at the heterostructure 
interface produces a natural coupling between the two $z$ valleys.  For an
infinite square well of width $L$, the Ohkawa-Uemura theory obtains a valley splitting of 
the form $E_v \sim \sin (2k_0L)/L^3$, where $k_0\hat{\bf z}$ is the location of the valley 
minimum in the Brillouin zone.\cite{ohkawa78}  This result was later confirmed by TB 
theory.\cite{boykin04,boykin04b}  The theory 
therefore captures the main qualitative aspects of the valley physics, with 
no additional input parameters besides those describing the bulk dispersion relation.  
In this sense, it is a first principles theory of valley splitting.

However, some aspects of the Ohkawa-Uemura theory have been called into question.
First, it has been criticized for its inaccurate description of the dispersion
relation near the bottom of the valleys,\cite{AFS} leading to quantitative 
errors.  More importantly, the method
relies on a closed EM description, which cannot easily incorporate 
microscopic details of the quantum well barrier.
This contradicts the fact that the valley coupling arises 
from physics occuring within several angstr\"{o}ms of the interface.\cite{sham79} 
Such distances are much smaller than any EM length scale, and cannot be accurately
described within any EM theory.

A physically appealing description of the heterostructure interface has been 
put forward by Sham and Nakayama.  
These authors develop a theory in which 
the reflection, transmission, and valley-scattering of waves at a Si/Si0$_2$ interface 
is built directly into the Bloch function basis states.  Since the confinement is 
incorporated into the basis set, it does not also appear as an external potential in 
the envelope equation.  Any additional potentials entering envelope 
equation (\textit{e.g.}, electrostatic potentials) are therefore smooth, 
and easily accommodated in an EM approach.  
The analytical results for the valley splitting are similar to Ohkawa-Uemura.  
Indeed, the two approaches have been shown to be closely related.\cite{ando79}

The Sham-Nakayama theory has also been criticized.  First, the theory is not 
self-contained -- a single input parameter $\alpha$ is introduced to characterize 
the microscopic width of the interface.  Although Sham and Nakayama
provide an estimate for $\alpha$, the parameter is phenomenological.  More 
importantly, the resulting EM theory is somewhat cumbersome, and cannot
provide simple analytical solutions for 
the heterostructure geometries considered here.

In this paper, we develop an EM theory which retains the desirable
qualities of both the Ohkawa-Uemura and Sham-Nakayama approaches.  We introduce a 
valley coupling parameter $v_v$, which efficiently describes the valley coupling 
for any type of interface, and which enables simple analytical results for the 
valley splitting that are in agreement with atomistic theories.

\section{Effective Mass Theory}
\label{sec:EMT}
The EM theory of Kohn and Luttinger \cite{kohn} provides an excellent description
of electrons in a semiconductor matrix under the influence of a
slowly varying confinement potential $V(\mathbf{r})$.\cite{daviesbook}  Here, 
``slowly" is defined with respect to the crystalline unit cell of length $a$:
\begin{equation}
V(\mathbf{r})/|{\bm \nabla} V(\mathbf{r})| \gg a .
\label{eq:criterion}
\end{equation}
In practice, the EM approach is extremely robust, often proving accurate 
well outside its range of validity.  Indeed, the standard textbook descriptions of shallow
donors and quantum wells are both based on an EM theory,\cite{daviesbook}
despite the singular nature of their confinement potentials.  

When the validity criterion (\ref{eq:criterion}) is not satisfied, it is a good
idea to compare the EM results with microscopic or atomistic approaches, such
as the TB theory of Sec.~\ref{sec:TB}.
For GaAs quantum wells, the EM theory provides quantitatively accurate results in most 
situations.  The approach only breaks down for very narrow quantum wells, or for high 
subband indices.\cite{long}  On the other hand, for indirect
gap semiconductors like silicon, the EM
theory must be extended if valley splitting becomes an important issue.  
A singular confinement potential causes valley coupling, and calls for a more 
sophisticated treatment.  Here, we provide a discussion of both the general
multi-valley EM approach, and of the valley coupling, which arises from a sharp 
quantum barrier.

In the standard EM theory, the wavefunction for a conduction electron in bulk Si can be written 
as a sum of contributions from the six degenerate valleys.  However, the lattice mismatch
between Si and Ge in a Si$_{1-x}$Ge$_x$/Si/Si$_{1-x}$Ge$_x$ quantum well causes tensile 
strain in the Si layer (assuming strain-relaxed SiGe).  As a result,
four of the six valleys rise in energy, while the two $z$ valleys
fall in energy.\cite{herring}  The strain splitting is on the order of 200~meV for 
the composition corresponding to $x=0.3$.\cite{schaeffler}  Consequently, only $z$ 
valleys play a role in typical low-temperature experiments.

\begin{figure}[t]
\centerline{\epsfxsize=3.in \epsfbox{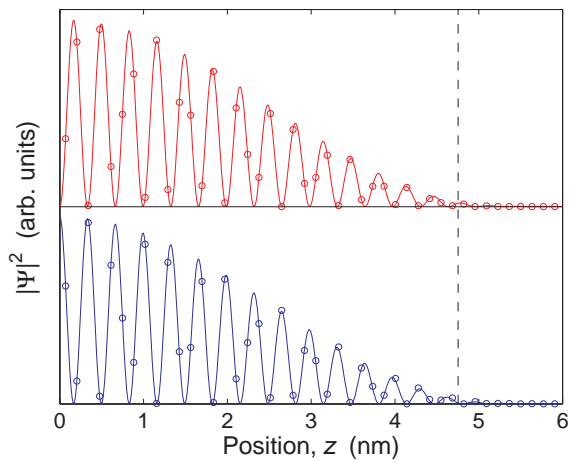}}
\caption{ 
(Color online.)
Comparison of effective mass and tight binding results for the two lowest eigenstates in a
Si$_{0.7}$Ge$_{0.3}$/Si/Si$_{0.7}$Ge$_{0.3}$ quantum well of width 9.5~nm.  (Only half 
the eigenfunctions are shown.)
The solutions correspond to the same orbital subband, but different valley states.
Bottom:  ground state.  Top:  excited state.  Solid lines:  effective mass theory.
Circles:  tight-binding theory.  Dashed line:  quantum well boundary.
\label{fig:envelope}}
\end{figure}

The EM wavefunction for strained silicon can then be expressed as
\begin{equation}
\Psi(\mathbf{r}) = \sum_{j=\pm z} \alpha_j e^{ik_jz} u_{\mathbf{k}_j}(\mathbf{r})
F_j(\mathbf{r}) . \label{eq:KL}
\end{equation}
The constants $\alpha_j$ describe the relative phase between the two $z$ valleys,
with $|\alpha_{+z}|=|\alpha_{-z}|=1/\sqrt{2}$.  
The functions $e^{ik_jz}u_{\mathbf{k}_j}(\mathbf{r})$ are Bloch functions, where
$\mathbf{k}_{\pm z}=\pm k_0\hat{\bm z}$ are the conduction valley minima.  We shall 
see that the envelope functions $F_j(\mathbf{r})$ are the same for the two $z$ valleys. 
 
A central feature of the EM formalism, which is a consequence of Eq.~(\ref{eq:criterion}), 
is that wavefunction separates neatly into atomic scale oscillations (the Bloch functions) 
and long wavelength modulations (the envelope function).
This is apparent in Fig.~\ref{fig:envelope}, where we show the EM and TB results for the
two valley states corresponding to the lowest subband of a quantum well, treating the Bloch
functions as described in Sec.~\ref{sec:TB}.  The two wavefunctions have
the same envelope, but their fast oscillations are phase shifted by $\pi/2$.
Note that although the EM and TB approaches are fundamentally different 
(discrete vs.\ continuous), their wavefunction solutions are almost identical.
Also note that the fast oscillations arising from $u_{\mathbf{k}_j}(\mathbf{r})$ (not 
pictured in Fig.~\ref{fig:envelope})
are commensurate with the crystal lattice, while the oscillations from $e^{ik_jz}$ 
are not, since $k_0$ is not at the Brillouin zone boundary.

We can draw two main conclusions from
the EM/TB comparison.  First, the EM treatment correctly captures the essence of
the subband physics, including both the long-wavelength and atomic scale 
features.  Higher subbands (not pictured in Fig.~\ref{fig:envelope}) are also 
described accurately.  Second, the main
difference between pairs of valley states does not occur in the envelope function, but
in the fast oscillations.  To leading order, the valley states are fully
characterized by their valley composition vectors ${\bm \alpha}=(\alpha_{-z},\alpha_{+z})$. 
Since the envelope function is independent of 
the valley physics at this order, the $\bm \alpha$ vectors may be obtained from 
first-order, degenerate perturbation theory, by treating the valley coupling as a
perturbation.\cite{koiller02}  
We now describe the perturbation theory, following the approach of 
Ref.~[\onlinecite{friesen05}], where shallow donors in silicon were considered.  
We specifically avoid the question of what causes valley
splitting, since this lies outside the scope of the EM theory.
Instead, we introduce valley coupling phenomenologically, through the parameter $v_v$
described below. 

In the EM formulation of Fritzsche and Twose,\cite{fritzsche,twose} the  
strained silicon wavefunction, Eq.~(\ref{eq:KL}), is determined from the equation
\begin{equation}
0=\sum_{j=\pm z} \alpha_j e^{-ik_jz} [H_0+V_v(z)-\epsilon ]F_j(z) . \label{eq:twose}
\end{equation}
Here, 
\begin{equation}
H_0 = T(z) +V_\text{QW}(z)+V_\phi(z) ,
\label{eq:EMH0}
\end{equation}
and
\begin{equation}
T=-\frac{\hbar^2}{2}\frac{\partial}{\partial z} \left(
\frac{1}{m_l} \, \frac{\partial}{\partial z} \right) \label{eq:T}
\end{equation}
is the one-dimensional kinetic energy operator.  
The longitudinal effective mass $m_l$ is materials dependent, and varies
from layer to layer in the heterostructure.  However, for Si-rich SiGe layers, $m_l$ depends
only weakly on the composition.  We therefore take $m_l\simeq 0.92\,m_0$ to be a constant.  
Note that the transverse effective mass $m_t$ does not appear in
Eq.~(\ref{eq:T}), since we initially consider only one-dimensional problems.  
In Sec.~\ref{sec:steps}, a more complicated, two-dimensional problem is studied, in
which $m_t$ appears.

Three different potentials energies appear in Eqs.~(\ref{eq:twose}) and (\ref{eq:EMH0}).  
$V_\text{QW}(z)$ describes the conduction band offsets in the heterostructure.  
For a quantum well, $V_\text{QW}(z)$ 
corresponds to a pair of step functions.  The 
electrostatic potential energy $V_\phi (z)$ describes any additional,
slowly varying potential.  Typically, $V_\phi (z)\sim -eEz$ is an electrostatic potential 
caused by modulation doping in the heterostructure or by external gates.  

The EM approximation breaks down near a singular confining potential 
like $V_\text{QW}(z)$, leading to a valley coupling.  Very near the singularity,
criterion (\ref{eq:criterion}) is not satisfied, and it becomes impossible to
fully separate the short-wavelength physics of the crystal matrix from the
long-wavelength confinement of the excess electron.  However, we may neatly capture the
valley interaction in terms of an effective coupling potential $V_v(z)$, which  
vanishes everywhere except within about an atomic length scale of 
the interface.  The detailed form of such a coupling may be quite 
complicated.\cite{pantelides}
However, because it is so strongly peaked, we may treat $V_v(z)$ as a $\delta$-function
over EM length scales.  Indeed, the $\delta$-function formulation arises 
naturally from some first principles theories of valley coupling at heterostructure 
interfaces.\cite{foreman05}  We then have
\begin{equation}
V_v(z)=v_v \delta (z-z_i) , \label{eq:Vv}
\end{equation}
where $z_i$ is the vertical position of the heterostructure interface.  
In Sec.~\ref{sec:steps}, we consider a case where $z_i$ depends on the lateral
position, $x$.  However, in the other sections
of the paper, we assume that $z_i$ is constant.
The valley interaction potential $V_v(z)$ plays a role analogous to the
central cell potential for an electron near a shallow donor.\cite{friesen05}  
The valley coupling strength $v_v$ is a
scalar quantity, which must be determined from experiments, 
or from atomistic methods like TB.

At lowest order in the perturbation theory (zeroth order),
we set $V_v(z)=0$ in Eq.~(\ref{eq:twose}).  Because of the fast oscillations associated with
the exponential factors, the contributions from the two valleys are approximately decoupled 
at this level, reducing Eq.~(\ref{eq:twose}) the conventional Kohn-Luttinger envelope equation:
\begin{eqnarray} && 
\left[ -\frac{\hbar^2}{2m_l} \, \frac{\partial^2}{\partial z^2} +V_\text{QW}(z)+V_\phi (z)
\right] F^{(0)}(z)
\\ && \hspace{2.05in}
= \epsilon^{(0)} F^{(0)}(z) . \hspace{.3in} \nonumber
\label{eq:envelope}
\end{eqnarray}
Here, the superscript $^{(0)}$ denotes an unperturbed eigenstate.  Note that the effective 
mass is the same for both $z$ valleys.  The corresponding envelopes are therefore
equivalent, and we shall drop the valley index.  

We now solve for $\bm \alpha$ and $\epsilon$ in Eq.~(\ref{eq:twose}) using 
first order perturbation
theory.  By replacing $F_j(z)$ in Eq.~(\ref{eq:twose}) with its zeroth order approximation,
left-multiplying by ${F^{(0)}}^*(z)\,e^{ik_lz}$, and integrating over $z$, we can express 
Eq.~(\ref{eq:twose}) in matrix form:
\begin{equation}
\begin{pmatrix}
\epsilon^{(0)}+\Delta_0&\Delta_1\\
\Delta_1^*&\epsilon^{(0)}+\Delta_0
\end{pmatrix}
\begin{pmatrix}
\alpha_{-z}\\
\alpha_{+z}
\end{pmatrix}=\epsilon
\begin{pmatrix}
\alpha_{-z}\\
\alpha_{+z}
\end{pmatrix}
. \label{eq:H}
\end{equation}
We have dropped small terms involving atomic scale oscillations in the integrand.
The perturbation terms are defined as follows:
\begin{eqnarray}
\Delta_0 &=& \int V_v(z) \left| F^{(0)}(z)\right|^2 \, dz , \label{eq:Delta0} \\
\Delta_1 &=& \int e^{-2ik_0z} V_v(z) \left|F^{(0)}(z)\right|^2 \, dz \label{eq:Delta1} .
\end{eqnarray}

Diagonalizing Eq.~(\ref{eq:H}) gives the first order energy eigenvalues
\begin{equation}
\epsilon_\pm = \epsilon^{(0)} +\Delta_0 \pm |\Delta_1| ,
\end{equation}
and the valley splitting
\begin{equation}
E_v=2|\Delta_1| .
\end{equation}
In the valley basis $(\alpha_{-z},\alpha_{+z})$, the eigenvectors 
corresponding to $\epsilon_\pm$ are given by
\begin{equation}
{\bm \alpha}_\pm = \frac{1}{\sqrt{2}} \left(
e^{i\theta/2} , \pm e^{-i\theta/2} \right) . \label{eq:evect}
\end{equation}
Here, $\pm$ refers to the even or odd valley combinations, and we have defined the 
phases
\begin{equation}
e^{i\theta} \equiv \Delta_1/|\Delta_1| . \label{eq:theta}
\end{equation}

For the case of a single interface at $z=z_i$, we obtain
\begin{equation}
\Delta_0=v_v|F^{(0)}(z_i)|^2 ,\hspace{.15in}
\Delta_1=v_ve^{-2ik_0z_i}|F^{(0)}(z_i)|^2 . \label{eq:DeltaA}
\end{equation}
We see that it is the magnitude of the envelope function at the interface that determines
the strength of the valley coupling.  The ${\bm \alpha}_\pm$ vectors for this geometry
are obtained from Eq.~(\ref{eq:evect}), with
\begin{equation}
\theta =2k_0z_i . \label{eq:thetaA}
\end{equation}

The valley coupling integrals in Eqs.~(\ref{eq:Delta0}) and (\ref{eq:Delta1}) provide a
simple but economical characterization of the valley splitting.
The coupling parameter $v_v$, which we compute below, 
provides a means to incorporate important microscopic details about the interface.  
The utility of the present theory is demonstrated by the ease with which we 
obtain results in the following sections.  The accuracy of the theory
is demonstrated in terms of\ the agreement between the EM and TB techniques.

We close this section by noting that several authors have
criticized the Fritzsche-Twose formulation of the multi-valley 
EM theory,\cite{shindo76,AFS} particularly because of the way intervalley
kinetic energy terms are calculated.\cite{ohkawa77a}  However, we emphasize that 
the present treatment
includes all first order corrections to the valley splitting.  Since the intervalley 
kinetic energy is of order $v_v^2$, it has a weaker contribution.  
The previous criticisms therefore do not apply here.

\section{Tight-Binding Theory}
\label{sec:TB}
We now discuss a tight binding method for modeling heterostructures in silicon.
Our main goal is to compare the solutions from such an atomistic technique with 
those of the EM theory, and to compute
the valley coupling parameter $v_v$, whose value cannot
be determined within the present EM theory.  We focus on the two-band TB model of 
Boykin \textit{et al.},\cite{boykin04,boykin04b} because of its simplicity.

In the two-band model,
the TB Hamiltonian includes nearest neighbor and next-nearest neighbor tunnel 
couplings, $v$ and $u$, respectively.  The values $v=0.683$~eV and 
$u=0.612$~eV are chosen such that (\textit{i}) the bulk dispersion relation 
$\epsilon (k)$ has two valleys, centered at $|k|=k_0=0.82\, (2\pi /a)$, and (\textit{ii}) 
the curvature of $\epsilon (k)$ at the bottom of a valley
gives the correct longitudinal effective mass $m_l=0.91\, m_0$.  
The unit cell in this theory consists of two atoms, with separation $a/4$ along 
the $[001]$ axis, where $a=5.431$~\AA~is the length of the silicon cubic unit 
cell.  These parameters correspond to bulk silicon. 
A more sophisticated theory should take into account compositional variations
and strain conditions.  However, for most situations of interest, the modified parameters
differ only slightly from the bulk.

In addition to the tunnel couplings $v$ and $u$, we can also include 
onsite parameters $\lambda_i$, to provide a locally varying confinement potential.  Both the 
conduction band offset $V_\text{QW}(z)$ and the electrostatic potential $V_\phi(z)$
can be expressed as on-site terms.  To avoid boundary errors,
we choose a TB lattice much larger than the confined electronic wavefunction.  Diagonalization
of the resulting TB Hamiltonian gives energy eigenvalues, from which we can calculate the
valley splitting.  

The two-band TB theory describes silicon valley physics with a minimal number (2) of input parameters, which are both fixed by fitting to measured band structure parameters.
More sophisticated techniques can
provide numerical improvements, but they generally do not capture any new 
physics.  In Ref.~[\onlinecite{boykin04}], a
comparison of the valley splitting between the two-band theory 
and a detailed many-band theory shows excellent qualitative agreement.  The more
accurate treatment gives results that are smaller by an approximately constant factor of 25\%.

\begin{figure}[t]
\centerline{\epsfxsize=3in \epsfbox{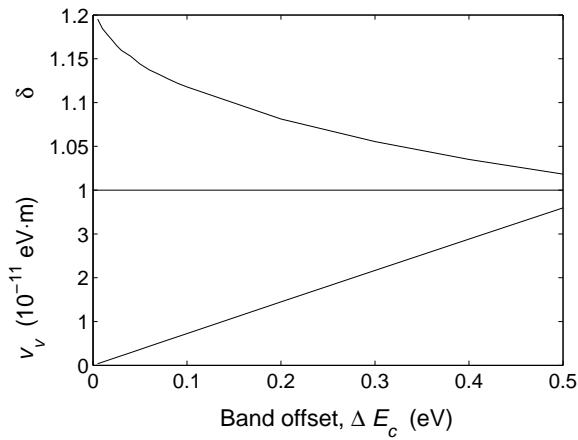}}
\caption{ 
The valley coupling parameter $v_v$ and the quantum well ``set-back"
parameter $\delta$, defined in
Eq.~(\ref{eq:deltaL}), as a function of the conduction band offset $\Delta E_c$.
The results are obtained for a symmetric, finite square well by comparing the EM and TB 
theories.  
\label{fig:TB}}
\end{figure}

Some typical TB eigenstates for a finite square well are shown in
Fig.~\ref{fig:envelope}.  The data points correspond to the squared TB amplitudes plotted at
the atomic sites.  Comparison with the full EM wavefunctions requires
knowledge of the Bloch functions in Eq.~(\ref{eq:KL}).  
However, to make contact with the TB 
results, we only need to evaluate the Bloch functions at the atomic sites.  
According to Eqs.~(\ref{eq:KL}) and (\ref{eq:squarealpha}) (see Sec.~\ref{sec:finite},
below), the low-lying valley pair of EM wavefunctions can be expressed as 
\begin{equation}
\Psi_\pm (z) = \frac{1}{\sqrt{2}} 
\left[ u_{-k_0}({\bm r})e^{-ik_0z} \pm u_{+k_0}({\bm r})e^{+ik_0z} \right] F(z) ,
\label{eq:fullBloch}
\end{equation}
where the ground-state alternates between $+$ and $-$ as a function of $L$.
We can denote the two atoms in the TB unit cell as $A$ and $B$, with the corresponding
Bloch functions $u_{k_0}(A)$ and $u_{k_0}(B)$.
By translational symmetry, we must have $u_{k_0}(A)=\pm u_{k_0}(B)$.  (Typically,
we observe the $-$ sign in our TB analyses.) 
The Bloch functions satisfy 
time-reversal symmetry, so that $u^*_{k_0}({\bm r})=u_{-k_0}({\bm r})$.  
Thus, defining $u_{k_0}(A)=|u_{k_0}(A)|e^{i\varphi}$ and assuming proper
normalization, we see that the squared TB amplitudes must fall on the curves
\begin{eqnarray}
|\Psi_+ (z)|^2 &=& 2\cos^2 (k_0z+\varphi )F^2(z) \label{eq:even} \\
|\Psi_- (z)|^2 &=& 2\sin^2 (k_0z+\varphi )F^2(z) \label{eq:odd} ,
\end{eqnarray}
where all information about the Bloch functions is reduced to the unimportant phase variable
$\varphi$.  Eqs.~(\ref{eq:even}) and (\ref{eq:odd}) are plotted in Fig.~\ref{fig:envelope}, 
setting $\varphi = 0$.  We see that these analytical forms provide an excellent representation
of the TB results.  However, we emphasize that Eqs.~(\ref{eq:even}) and (\ref{eq:odd})
are accurate only at the atomic sites.  To describe the
wavefunction between the atomic sites would require additional knowledge of the silicon 
Bloch functions.\cite{hu05,wellard03,wellard05,hollenberg05} 

We can use the TB theory to determine the EM valley coupling parameter $v_v$ 
by comparing corresponding results in the two theories.  This is accomplished in 
Sec.~\ref{sec:finite} for
the finite square well geometry, with results shown in Fig.~\ref{fig:TB}.

\section{Finite Square Well}
\label{sec:finite}
We consider a symmetric square well with barrier interfaces
at $z_i=\pm L/2$, corresponding to a quantum well of width $L$.  
We assume that the two interfaces are equivalent, so the same valley coupling $v_v$ 
can be used on both sides.  
Using EM theory, the resulting valley splitting is
\begin{equation}
E_v = 4 v_v F^2(L/2) |\cos (k_0L)| .  \label{eq:FSWEv}
\end{equation}

\begin{figure}[t]
\centerline{\epsfxsize=3.4in \epsfbox{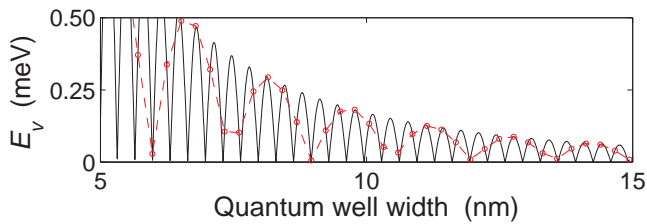}}
\caption{ 
(Color online.)
Valley splitting in a finite square well.
We compare effective mass results (solid curve) with  tight binding 
results (circles), as a function of the quantum well widths
$L$ and $L'$, respectively (see text).
The TB data points occur at integer multiples of the TB unit cell.  We assume a
Si$_{0.7}$Ge$_{0.3}$/Si/Si$_{0.7}$Ge$_{0.3}$ quantum well, corresponding to
$\Delta E_c = 160$~meV.
\label{fig:QW}}
\end{figure}

An analytical solution of the EM equations for the envelope function of a finite square 
well can be obtained 
by matching wavefunction solutions at the interfaces,\cite{daviesbook}  giving
\begin{equation}
F(L/2) = \left[ \frac{1}{k_b}+\frac{k_wL+\sin (k_wL)}{2 k_w\cos^2 (k_wL/2)} \right]^{-1/2} ,
\label{eq:f0QW}
\end{equation}
where
\begin{equation}
k_b=k_w \tan (k_wL/2) . \label{eq:kbkw}
\end{equation}
The wavevector $k_w$ can be obtained numerically from the 
transcendental equation
\begin{equation}
k_w^2 =\frac{2m_l\Delta E_c}{\hbar^2} \cos^2 (k_wL/2) . \label{eq:transcend}
\end{equation}
Some typical results for the valley splitting as a function of the well width are shown in 
Fig.~\ref{fig:QW}. 

An approximate solution for Eqs.~(\ref{eq:FSWEv})-(\ref{eq:transcend}) can be obtained in the 
limit of a very deep or a very wide quantum well.  When 
$\pi^2\hbar^2/2m_l L^2 \ll \Delta E_c$, we find that $k_w \simeq \pi/L$ for the first subband,
leading to
\begin{equation}
F^2(L/2) \simeq \frac{\pi^2 \hbar^2}{m_l \Delta E_c L^3} .
\end{equation}
The valley splitting is then given by
\begin{equation}
E_v \simeq \frac{4 v_v \pi^2 \hbar^2}{m_l \Delta E_c L^3} |\cos (k_0 L)| .
\label{eq:EvB}
\end{equation}
This agrees with the dependence on well width obtained in Refs.~[\onlinecite{boykin04}] and
[\onlinecite{ohkawa78}], for an infinite square well.

Diagonalization of the perturbation Hamiltonian in Eq.~(\ref{eq:H}) gives the ground 
($g$) and excited ($x$) valley state $\bm \alpha$ vectors
\begin{equation}
\begin{array}{ll}
{\bm \alpha}_g = (1,-\text{sign} [ \cos (k_0L) ] )/\sqrt{2} & \quad \text{(ground)} , \\
{\bm \alpha}_x = (1,\text{sign} [ \cos (k_0L) ] )/\sqrt{2} & \quad \text{(excited)} ,
\end{array} \label{eq:squarealpha}
\end{equation}
where we have defined $\text{sign}[x]\equiv x/|x|$.
These results are equivalent to Eq.~(\ref{eq:evect}), up to an overall phase factor.
We see that the oscillations in Fig.~\ref{fig:QW} correspond to alternating
even and odd ground states [${\bm \alpha}_+=(1,1)/\sqrt{2}$ and 
${\bm \alpha}_-=(1,-1)/\sqrt{2}$, respectively], as a function of
of $L$.  This alternating behavior has been observed previously in  
TB analyses.\cite{boykin04} 

We can use these results to obtain an estimate for the valley coupling parameter
$v_v$, by comparing the EM and TB results for the finite square well.  For the EM case,
we first solve Eq.~(\ref{eq:transcend}) numerically,
to obtain $k_w$.  We then solve Eq.~(\ref{eq:f0QW}) for $F(L/2)$,
finally obtaining the valley splitting from Eq.~(\ref{eq:FSWEv}).
For a given value of $\Delta E_c$, we fit Eq.~(\ref{eq:FSWEv}) to the 
numerical TB results, as a function of $L$, using $v_v$ as a fitting parameter.  
However, there is an ambiguity in relating the quantum well width, $L$, in the continuous
EM theory to the discretized width, $L'=Na/2$, in the TB theory, where $N$ is the
number of silicon TB unit cells in the quantum well.  For example, is the interface located on
the last atomic site in the quantum well, the first atomic site in the barrier region, 
or somewhere in between?  We see that 
the two widths, $L$ and $L'$, may differ on the scale of a single atomic layer, or
$a/4=1.36$~\AA.
To allow for this, we introduce a second fitting parameter $\delta$, as defined by 
\begin{equation}
L=L'+\delta \,a/4 . \label{eq:deltaL}
\end{equation}  
With these two fitting parameters, $v_v$ and $\delta$,
we obtain nearly perfect correspondence between the EM and TB theories, 
as shown in Fig.~\ref{fig:QW}.  

In this manner, we can map out $v_v$ and $\delta$ as a function of $\Delta E_c$, giving the
results shown in Fig.~\ref{fig:TB}.  We find that $v_v(\Delta E_c)$ is 
linear over its entire range, with 
\begin{equation}
v_v = 7.2\times 10^{-11}\Delta E_c . \label{eq:vvlinear}
\end{equation}
Here, $v_v$ is given in units of eV$\cdot$m when $\Delta E_c$ is expressed in units
of eV.  As described in Sec.~\ref{sec:TB}, a many-band TB analysis obtains results 
for $v_v$ that are smaller by a factor of about 25\%.

From Fig.~\ref{fig:TB}, we see that $\delta \simeq 1.1$ forms a reasonable approximation 
over the typical experimental range, $\Delta E_c\simeq 50$-200~meV.  
This corresponds to about one atomic 
layer, or half an atomic layer on either side of the quantum well.  We can
interpret $\delta$ as the set-back distance for an effective scattering barrier
which causes  valley
coupling.  A similar interpretation was given for the parameter $\alpha$ in 
Ref.~[\onlinecite{sham79}].  We see that this set-back distance increases for a shallow 
quantum well.  
 
Finally, we consider the asymptotic limits of our EM theory. 
In the limit $\Delta E_c\rightarrow \infty$, corresponding to an infinite square well, 
Eq.~(\ref{eq:EvB}) becomes exact.  Since
$E_v$ does not vanish, we conclude that $v_v\propto \Delta E_c$ in this limit.
This is precisely the behavior observed in Fig.~\ref{fig:TB}.

In the limit $\Delta E_c\rightarrow 0$, corresponding to a shallow square well,
Eq.~(\ref{eq:EvB}) is not valid.  Instead, we obtain
$F^2(L/2)\simeq \Delta E_cLm_l/\hbar^2$, leading to
$E_v\simeq 2\Delta E_cLv_vm_l\cos (k_0L)/\hbar^2$.  In this limit, we expect the 
valley splitting to vanish, but we can make no other predictions about $v_v$.  
The numerical results, however, suggest that the 
linear dependence of $v_v(\Delta E_c)$ extends smoothly to zero.

\section{Quantum Well in an Electric Field}
\label{sec:Efield}
\begin{figure}[t]
\centerline{\epsfxsize=2.5in \epsfbox{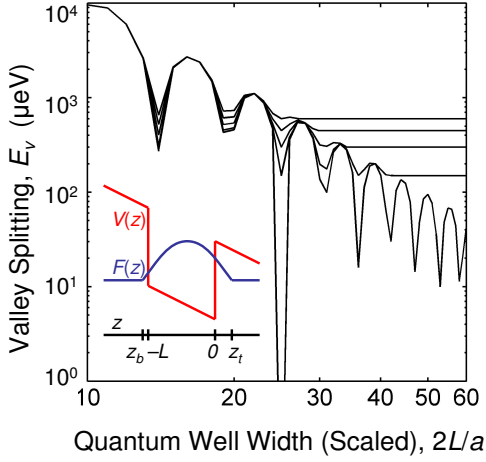}}
\caption{ 
(Color online.)
Effective mass results for the valley splitting in an electric field, as a function of the 
quantum well width.  Five $E$ fields are shown, from bottom to top:  
$E=0, 1.04, 2.08, 3.12$, and 4.16~MV/m.
The well width $L$ is scaled by the TB unit cell size, $a/2=2.716$~\AA.
Results are shown for a Si$_{0.8}$Ge$_{0.2}$/Si/Si$_{0.8}$Ge$_{0.2}$ quantum well,
analogous to Fig.~1(a) of Ref.~[\onlinecite{boykin04}].
Inset:  $E$-field geometry, with confinement potential $V(z)=V_\text{QW}(z)+V_\phi(z)$ and 
variational wavefunction $F(z)$ given in Eq.~(\ref{eq:Eform}).
Here, $z_b=z_t-\pi/k$.
\label{fig:Efield}}
\end{figure}

We now consider a quantum well in the presence of an electric field oriented 
in the growth direction.  
The geometry is shown in the inset of Fig.~\ref{fig:Efield}.
For physically realistic fields, caused by modulation doping or electrical top-gates,
the resulting electrostatic potential satisfies the EM criterion, Eq.~(\ref{eq:criterion}).
We therefore proceed as in Sec.~\ref{sec:finite}, using the
numerical values for $v_v(\Delta E_c)$, obtained for a symmetric square well.
The electrostatic potential enters our analysis through the envelope 
equation (\ref{eq:envelope}).  Although there are no exact solutions for the problem 
of a tilted square well, the approximate treatment described here provides analytic 
results that accurately reproduce the results of TB theory. 

We assume an electrostatic potential energy 
given by $V_\phi(z)=-eEz$, and a quantum well of
width $L$ and height $\Delta E_c$.  The top barrier of the well lies 
at $z=0$.  We consider the following variational envelope function:
\begin{equation}
F(z)= \left\{
\begin{array}{ll}
-\sqrt{\frac{2k}{\pi}} \sin [k(z-z_t)] &\quad \left( z_t-\frac{\pi}{k} <z<z_t \right) \\
0 & \quad \text{(otherwise)} .  
\end{array} \right. \label{eq:Eform}
\end{equation}
For simplicity, we have chosen a wavefunction tail that terminates abruptly.  
This approximation is satisfactory for a variational calculation, since the tail, which
is exponentially suppressed, contributes very little to the energy expectation value.  

In our trial function, the parameter $k$ accounts for the finite barrier height by allowing 
the wavefunction to extend into the barrier region.  The upward shift 
of the wavefunction in the presence of an electric field is given by $z_t$.  
The solution (\ref{eq:Eform}) becomes exact for an infinite square well at zero field, suggesting that the trial function will be most effective in this 
limit.  We therefore define the small parameters
\begin{equation}
x = \frac{\hbar^2\pi^2}{2m_lL^2\Delta E_c} \quad \text{and} \quad
y = \frac{eEL}{4\Delta E_c}.
\end{equation}

The energy expectation value, obtained from envelope equation (\ref{eq:KL}), can be 
expressed in terms of the dimensionless variational parameters
\begin{equation}
\theta_1=kL \quad \text{and} \quad \theta_2=k z_t ,
\end{equation}
giving
\begin{widetext}
\begin{equation}
\frac{\pi^2 \epsilon}{\Delta E_c} \simeq \left\{
\begin{array}{ll}
x\theta_1^2-4\pi^2y(\theta_2-\pi/2)/\theta_1
+2\pi[\theta_2^3-(\theta_1+\theta_2-\pi)^3]/3  & \quad (\theta_1+\theta_2<\pi) \\
x\theta_1^2-4\pi^2y(\theta_2-\pi/2)/\theta_1
+2\pi\theta_2^3/3  & \quad (\theta_1+\theta_2\ge \pi) .
\end{array} \right.
\label{eq:Etheta}
\end{equation}
\end{widetext}
In the first case in Eq.~(\ref{eq:Etheta}), the wavefunction extends past both barriers.  
For large enough electric fields ($y>x$), the wavefunction only extends past one barrier 
(see inset of Fig. 4). 
Minimization of $\epsilon$ with respect to $\theta_1$ and $\theta_2$ gives
\begin{eqnarray}
\left. \begin{array}{l}
\theta_1 \simeq \pi+\sqrt{x-y}-\sqrt{x+y} \\
\theta_2 \simeq \sqrt{x+y} 
\end{array} \right\} & \quad (x>y), \\
\left. \begin{array}{l}
\theta_1 \simeq \pi (y/x)^{1/3} \\
\theta_2 \simeq (8y^2x)^{1/6} 
\end{array} \right\} \quad \quad \quad & \quad (x\le y) ,
\end{eqnarray}
where we have made use of the fact that $\theta_2 \ll \pi$.  For the case $x>y$, we 
have also used $\pi-\theta_1 \ll \pi$.

To compute the valley splitting, we use the results 
$F^2(-L)\simeq 2(\pi-\theta_1-\theta_2)^2/L$ and $F^2(0)\simeq 2\theta_2^2/L$ when $x>y$,
and $F^2(0)\simeq 2\theta_2^2\theta_1/\pi L$ when $x\le y$ to obtain
\begin{equation}
E_vL \simeq \left\{
\begin{array}{ll}
4v_v \left| e^{ik_0L}(x-y)+e^{-ik_0L}(x+y) \right| & \quad (x>y) \\
8v_vy & \quad (x\le y) .
\end{array} \right. \label{eq:EvwithE}
\end{equation}
These solutions are plotted in Fig.~\ref{fig:Efield} as a function of the quantum well width 
for several different $E$ fields.  In the figure, we have evaluated Eq.~(\ref{eq:EvwithE}) 
only at integer multiples of the TB unit cell, to facilitate comparison with Fig.~1(a) in
Ref.~[\onlinecite{boykin04}].  The correspondence between the EM and TB theories
is quantitatively and qualitatively accurate, particularly for large well widths.  

In Fig.~\ref{fig:Efield}, we see that the crossover to high field behavior 
corresponds to $E_v$ becoming independent of $L$.  The crossover occurs when $y>x$ or
\begin{equation}
E>\frac{2\pi^2\hbar^2}{m_leL^3} .
\end{equation}
At low fields, the valley splitting
exhibits interference oscillations, similar to Fig.~\ref{fig:QW}.
At high fields, the envelope function no longer
penetrates the barrier on the bottom side of the quantum well.  The top and 
bottom barriers then no longer produce interfering contributions to the valley splitting,
causing the oscillations in $E_v(L)$ to cease.  Since we have chosen a trial wavefunction
with no tail, the crossover to high field behavior in Fig.~\ref{fig:Efield}
occurs abruptly.  This is in contrast with the TB results where
small oscillations can still be observed right above the crossover.

We now study the asymptotic behaviors of Eq.~(\ref{eq:EvwithE}).
In the zero field limit, $y\rightarrow 0$, Eq.~(\ref{eq:EvwithE}) correctly reduces 
to Eq.~(\ref{eq:EvB}) for a symmetric square well.  In the high field limit,
we can ignore the bottom barrier entirely.  For an infinite barrier,
the problem is often analyzed using the Fang-Howard trial 
wavefunction.\cite{daviesbook}  To make contact with this approach, we shall now
perform a modified Fang-Howard analysis, and demonstrate a correspondence between the 
two results.

The conventional Fang-Howard trial function does not penetrate the barrier region.
However, for a finite barrier height $\Delta E_c$, it is the
amplitude of the wavefunction at the interface which determines the valley splitting.
We must therefore modify the Fang-Howard trial function to allow the wavefunction
to penetrate the top barrier, similar to Eq.~(\ref{eq:Eform}).  
An appropriately modified trial function is given by
\begin{equation}
F(z)= \left\{ 
\begin{array}{ll} 
-\sqrt{b^3/2}(z-z_t) \exp [b(z-z_t)/2] & \text{ for }z<z_t \\
0 & \text{ for }z\geq z_t 
\end{array} \right. . \label{eq:FH}
\end{equation}
This is just the usual Fang-Howard variational function, shifted upward by 
$z_t$.  In the Fang-Howard approach, the wavefunction tail decays exponentially 
into the lower portion of the quantum well.  This treatment is 
more physical than the abrupt termination assumed in Eq.~(\ref{eq:Eform}).  However, the
tail contribution to the variational calculation is insignificant, to leading order.  
Following the conventional Fang-Howard approach, but now using
$b$ and $z_t$ as variational parameters, we obtain 
\begin{equation}
E_v = \frac{2v_v eE}{\Delta E_c} . \label{eq:EvFH}
\end{equation}
As in the case of a symmetric square well, we see that $E_v \propto \Delta E_c^{-1}$.
We also find that Eq.~(\ref{eq:EvFH}) is equivalent to the high-field limit of
Eq.~(\ref{eq:EvwithE}).  Thus, the non-oscillating, high-field portions of the curves
in Fig.~\ref{fig:Efield} correspond to the ``Fang-Howard limit" described by
Eqs.~(\ref{eq:FH}) and (\ref{eq:EvFH}).  This agreement demonstrates that the variational form
of Eq.~(\ref{eq:Eform}) is robust over the entire field range.

\section{Valley Splitting in a 2DEG}
\label{sec:2DEG}
In this section and the next, we consider problems of particular experimental
importance.  In both cases, conventional techniques like TB theory are
somewhat cumbersome.  However, the EM formalism leads to straightforward solutions.
We first consider a two-dimensional electron gas (2DEG) 
in a SiGe/Si/SiGe quantum well, which extends the single electron senarios we
have studied so far, by including many-body interactions.  

\begin{figure}[t]
\centerline{\epsfxsize=2.5in \epsfbox{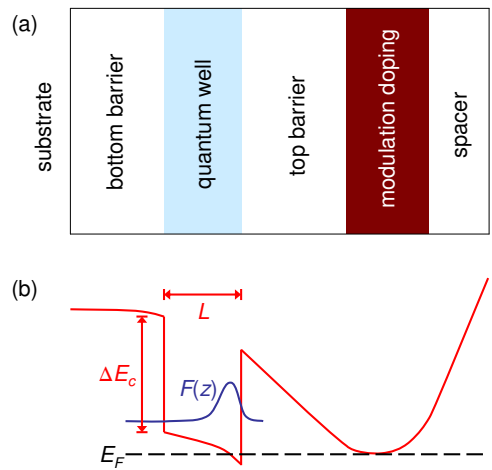}}
\caption{ 
(Color online.)
Typical 2DEG structure for a SiGe/Si/SiGe quantum well.
(a)  Heterostructure, left to right:  strain-relaxed SiGe substrate and barrier,
strained silicon quantum well, strain-relaxed SiGe barrier, $n^+$ SiGe doping layer,
and SiGe spacer layer.
(b)  Conduction band profile, showing the envelope function $F(z)$ and the Fermi energy $E_F$.
\label{fig:2DEG}}
\end{figure}

We consider the modulation-doped heterostructure shown in Fig.~\ref{fig:2DEG}(a).   
In this structure, we assume that the charge is found only in the 2DEG and the doping 
layers.  A more detailed analysis could also include background charge and charge trapped at
an interface.  Because 
the modulation doping field is large in a typical heterostructure, we will 
ignore the bottom barrier in our calculations.  The conduction band profile is sketched in 
Fig.~\ref{fig:2DEG}(b).  We treat many-body interactions using the 
Hartree approximation, as common for a quantum well.  However,
other many-body interactions can also be included,\cite{AFS,paul} using similar techniques
to calculate the valley splitting.  

The modified Fang-Howard approach gives a reasonable approximation 
for an electron in a 2DEG.\cite{AFS}  However, we must include the electron-electron 
interactions self-consistently.  Within the Hartree 
approximation,\cite{daviesbook} the 2DEG charge density $\rho (z)$ is given by
\begin{equation}
\rho(z)=-enF^2(z) ,
\end{equation}
where $n$ is the density of the 2DEG and $F(z)$ is defined in Eq.~(\ref{eq:FH}).
The charge density is related to
the electrostatic potential through the Poisson equation,
\begin{equation}
\frac{d^2\phi_H}{dz^2}=-\frac{\rho}{\varepsilon}, \label{eq:poisson}
\end{equation}
where $\varepsilon$ is the dielectric constant of silicon.
The boundary conditions on Eq.~(\ref{eq:poisson}) are given by
\begin{equation}
\begin{array}{ll}
d\phi_H/dz = 0 & \quad (z\rightarrow -\infty) ,  \\
\phi_H = 0 & \quad (z=0) . 
\end{array} \label{eq:BC}
\end{equation}
The second boundary condition anchors the energy of the confinement potential, 
$V_\text{QW}(z)$, at top of the quantum well.\cite{daviesbook}  
The corresponding electrostatic potential is given by
\begin{eqnarray}
\phi_H (z) &=& \frac{en}{2b \varepsilon} \left\{
[b^2(z-z_t)^2-4b(z-z_t)+6]e^{b(z-z_t)} 
\right. \nonumber \\ 
&& \hspace{.5in} \left. -[b^2z_t^2+4bz_t+6]e^{-b z_t} \right\}
\end{eqnarray}

The variational parameters are determined by minimizing the total energy per electron,
given by
\begin{equation}
\epsilon = \langle T \rangle + \frac{1}{2} \langle V_\phi \rangle
+\langle V_\text{QW} \rangle . \label{eq:Hartree}
\end{equation}
Here, $V_\phi(z)$ corresponds to the Hartree potential $-e\phi_H (z)$.  The factor of $1/2$ 
in the Hartree term prevents overcounting of the interactions.\cite{daviesbook}
Note that, in contrast with calculations for Si/SiO$_2$ inversion layers, we do not
include any contributions from image potentials in Eq.~(\ref{eq:Hartree}), since the
dielectric constants for Si and SiGe are nearly equal, and any other 
interfaces that could produce images are far away from the 2DEG.  

We evaluate Eq.~(\ref{eq:Hartree}), obtaining
\begin{equation}
\epsilon \simeq \frac{\hbar^2 b^2}{8m_l} +\frac{e^2n}{4b\varepsilon}
\left( \frac{33}{8}-2bz_t \right)+\frac{\Delta E_c \, b^3z_t^3}{6} ,
\end{equation}
where we have made use of the dimensionless small parameters $bz_t$ and 
$e^2n/\varepsilon \Delta E_cb$.
The first of these describes the shift of the wavefunction towards $+\hat{\bm z}$,
while the second describes the relative magnitude of the electrostatic energy with respect
to the band offset $\Delta E_c$.
Minimization of $\epsilon$ with respect to $b$ and $z_t$ gives
\begin{equation}
z_t^2\simeq \frac{8\hbar^2}{33m_l\Delta E_c} 
\hspace{.2in}\text{and}\hspace{.2in}
b^3\simeq \frac{33e^2nm_l}{8\hbar^2} .
\end{equation}

We can now calculate the valley splitting for a 2DEG.  Under the previous approximations,
we obtain a very simple expression for the wavefunction at the top quantum well interface,
\begin{equation}
F^2(0)\simeq \frac{e^2 }{2\varepsilon \Delta E_c} \label{eq:F2QW},
\end{equation}
leading to the valley splitting
\begin{equation}
E_v = \frac{v_v e^2n}{\varepsilon \Delta E_c} .  \label{eq:Ev2DEG}
\end{equation}
[Note that this result is obtained for a perfectly smooth interface.  
As we show in Sec.~\ref{sec:steps}, substrate roughness can reduce this estimate 
considerably.  However, the scaling dependence in Eq.~(\ref{eq:Ev2DEG}) remains valid.]

We can use our estimate for the valley coupling parameter, Eq.~(\ref{eq:vvlinear}), 
to obtain a quantitative prediction for the valley splitting in a 2DEG.  
Expressing the valley splitting in
units of meV and the 2DEG density $n$ in units of $10^{12}\,\text{cm}^{-2}$, we find that
\begin{equation}
E_v\simeq 1.14\, n . \label{eq:2DEGest}
\end{equation} 
Here, we have used the low-temperature dielectric constant for silicon, 
$\varepsilon = 11.4 \varepsilon_0$.  It is interesting to note that the barrier height 
$\Delta E_c$ does not directly enter the final result.  

We can compare Eq.~(\ref{eq:2DEGest})
with the corresponding, non-self-consistent calculation, by treating the system  
of 2DEG and doping layer
as a parallel plate capacitor with electric field $E=en/\varepsilon$.  The valley
splitting for an electron in such a field is given in Eq.~(\ref{eq:EvFH}), with the result
\begin{equation}
E_v\simeq 2.29\, n \hspace{.5in} \text{(non-self-consistent)}. \label{eq:2DEGestb}
\end{equation} 
The factor of 2 difference with Eq.~(\ref{eq:2DEGest}) arises because the electric field
in a real 2DEG is not uniform, due to the presence of charge.  The non-self-consistent procedure
therefore uses an electric field that is too large, overall, and it overestimates the
valley splitting.

In Eq.~(\ref{eq:2DEGest}), the prefactor 1.14
can be compared with similar estimates for silicon inversion layers.  Ohkawa and Uemura
obtain a prefactor of 0.15,\cite{ohkawa77a,ohkawa77b} 
while Sham and Nakayama obtain 0.33.\cite{nakayama78} 
A recent experiment in a top-gated Si0$_2$/Si/Si0$_2$ heterostructure has obtained
a much larger value of the valley splitting.\cite{takashina06}  However, several remarks
are in order.  First, we do not specifically consider top-gated structures here, so
Eq.~(\ref{eq:2DEGest}) does not directly apply to the latter experiment.  
Second, we point out that 
depletion and image charges, which were not considered here, play a more significant role 
in an inversion layer than a quantum well geometry.  Finally, we note that a more 
accurate, many-band estimate for $v_v$ would reduce the prefactor in
Eq.~(\ref{eq:2DEGest}) by about 25\%, as discussed in Sec.~\ref{sec:TB}.

\begin{figure}[t]
\begin{center}
\includegraphics[width=2.1in]{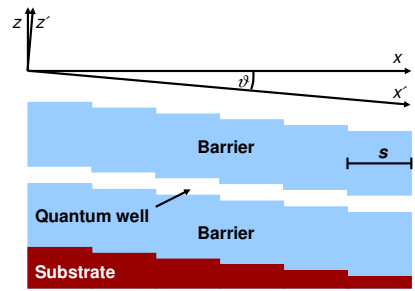}
\caption{
(Color online.)
Tilted quantum well geometry, with 
crystallographic axes $(x,y,z)$ and rotated axes $(x',y',z')$, assuming $y=y'$.
The effective tilt angle is $\vartheta$, and the atomic step size is $s$.
\label{fig:oscillations}}
\end{center}
\end{figure}

\section{Tilted Quantum Well in a Magnetic Field}
\label{sec:steps}
In high-mobility Si/SiGe devices, substrates are often
intentionally tilted away from $[001]$ by up to several degrees, to promote uniform 
epitaxial growth under strain conditions, and to help reduce step bunching.  
Additional roughening occurs during the growth of strained heterostructures.  
(Recent advances in nanomembrane technology may help overcome this 
difficulty.\cite{michele})
Typical devices therefore contain atomic steps at their 
surfaces and interfaces, associated with global or local tilting.  At the 
EM level, one might expect to ignore such small, atomic-scale steps, and to
work in the locally tilted basis $(x',y',z')$ shown in Fig.~\ref{fig:oscillations}.
However, when valley coupling is taken into account, the problem becomes
two-dimensional, since the tilted surface is misaligned with respect to the 
crystallographic axes.  In such high
dimensional geometries, there is an obvious scaling difficulty for atomistic theories.
However, the EM theory of valley splitting has a definite, practical advantage.  
The following discussion expands upon our previous analysis in 
Ref.~[\onlinecite{friesen06}].

We consider a tilted quantum well in a magnetic field.  In the 
low-field limit, the electronic wavefunction covers many steps, leading to interference
effects in Eq.~(\ref{eq:Delta1}), and a near-total suppression of the valley 
splitting.  To see this, we note that the vertical position of the interface,
$z_i$, becomes a function of the lateral position, $x$.
The phase factor in Eq.~(\ref{eq:Delta1}) is therefore not a constant, in contrast with
the case of a flat interface.  In Fig.~\ref{fig:oscillations},
we assume that a single step corresponds to a change of one atomic plane, or
$a/4=1.36$~\AA.  From Eq.~(\ref{eq:Delta1}), 
the phase shift between neighboring steps is given by
$2 k_0 (a/4) = 0.82 \, \pi$.  So the steps are nearly $180^\circ$ out
of phase, and the interference is severe.

At low magnetic fields, the experimental data are consistent with this picture of suppressed
valley splitting.\cite{goswami06}  (The experiments also suggest a non-vanishing 
zero-field extrapolation, which we shall investigate elsewhere.\cite{friescop})  
Some of the characteristic features of the magnetic field dependence can be explained 
in terms of the lateral confinement of the electronic wavefunction.
Magnetic confinement over the length scale
$l_B=\sqrt{\hbar/|eB|}$ reduces the number of steps that contribute to the valley
splitting integral.  The interference effects and the suppression of 
the valley splitting are similarly reduced.  
To get a sense of the scales involved, we note that a wavefunction of 
width $4l_B$ will cover $16(l_B/a)\tan \vartheta$ steps, where $\vartheta$ is the 
local tilt angle in Fig.~\ref{fig:oscillations}.  (Here, we have assumed a wavefunction of
diameter $2\sqrt{\langle r^2 \rangle}=4l_B$, obtained using the solutions
described below.)  For a typical $\vartheta =2^\circ$ miscut,\cite{goswami06}
the wavefunction covers about 26 steps when $B=1$~T, and 13 steps when $B=4$~T.  
The confinement provided by 
electrostatic top-gates can also enhance the valley splitting by reducing the
step coverage.\cite{goswami06}

We can gain insight into the magnetic field dependence of the valley splitting
by considering a simple model.
We first express the envelope function equation in the tilted basis $(x',y',z')$, giving
\begin{eqnarray}  & & \hspace{-.2in}
\Biggl[ \sum_{n=1}^3 \frac{1}{2m_n} \left( -i\hbar \frac{\partial}{\partial x'_n} 
+e A_n({\bm r'}) \right)^2 \Biggr. 
\label{eq:schrorot}
\\ & & \hspace{0.9in} \Biggl. 
+ V_\text{QW}(z')  \Biggr] F({\bm r'}) = \epsilon
\, F({\bm r'}) , \nonumber
\end{eqnarray}
where $m_1=m_2=m_t=0.19\, m_0$ and $m_3=m_l=0.92\, m_0$, are the transverse and lateral 
effective masses, respectively.  We note that the anisotropic effective masses are 
defined with respect to the crystallographic axes $(x,y,z)$, not the growth axes.
The full effective mass tensor in Eq.~(\ref{eq:schrorot}) should therefore include
off-diagonal terms.  In particular, we should have a term proportional to
$m_{xz}^{-1}\simeq \vartheta (m_t^{-1}-m_l^{-1})$, where we have taken 
$\vartheta \ll \pi/2$.  For a $2^\circ$ miscut, however, we find 
that $m_t/m_{xz}=0.028$, so the off-diagonal term is much smaller than the diagonal terms.
If desired, the off-diagonal corrections could be included, perturbatively.  Here, we 
have considered only the leading order (diagonal) mass terms.

In Eq.~(\ref{eq:schrorot}), the magnetic field is introduced through the vector potential.
We consider the symmetric gauge, with ${\bf A}({\bm r}')=(-y',x',0)B/2$.  We also 
assume an approximate form for the quantum well potential $V_\text{QW}(z')$, which
is smoothly tilted (\textit{i.e.}, not step-like).  This 
will be adequate for our simple estimate.
Separation of variables then leads to solutions of form $F({\bm r}')=F_{xy}(x',y')F_z(z')$, 
where $F_z(z')$ is the quantum well wavefunction, studied elsewhere in this
paper, and $F_{xy}(x',y')$ is the lateral wavefunction given by\cite{daviesbook}
\begin{eqnarray}
F_{xy}^{(nl)}(r',\theta') 
&=&  \sqrt{\frac{n!}{\pi l_B^2 2^{|l|+1}(n+|l|)!}} \,
e^{il\theta'}  
\label{eq:LL}  \\ && \hspace{.1in} \times
e^{-{r'}^2/4l_B^2} \left(\frac{r'}{l_B}\right)^{|l|}
L_n^{(|l|)}\left(\frac{{r'}^2}{2l_B^2} \right) . \nonumber 
\end{eqnarray}
Here, we use radial coordinates, defined as $(x',y')=(r'\cos \theta',r'\cos \theta')$, while
$n=0,1,2,\dots$ are the radial quantum numbers (the Landau level indices), 
$l=0,\pm 1, \pm 2, \dots$ are the azimuthal quantum numbers, and $L_n^{(|l|)}(x)$
are associated Laguerre polynomials.\cite{stegun}  
The energy eigenvalues for $F_{xy}^{(nl)}(r',\theta')$ are given by
\begin{equation}
\epsilon_{nl}=\frac{2n+l+|l|+1}{2} \, \hbar \omega_c,
\end{equation}
where $\omega_c=e|B|/m_t$ is the cyclotron frequency.  Note that we have ignored spin
physics here.

\begin{figure}[t]
\begin{center}
\includegraphics[width=2.1in]{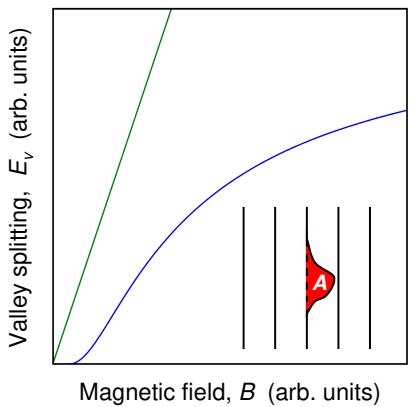}
\caption{
(Color online.)
Valley splitting on a tilted quantum well, in a magnetic field.
The lower (blue) curve shows the exponential suppression of the valley splitting, 
Eq.~(\ref{eq:EvsB}), arising from
perfectly uniform steps.  The upper (green) curve shows the linear behavior arising from the
``plateau" model of disordered steps.  Inset shows a localized step fluctuation 
(or ``wiggle") of area $A$, as analyzed in the text.
\label{fig:EvsB}}
\end{center}
\end{figure}

To take an example, we now focus on the lowest Landau level, with $n=l=0$.  The behavior
of the valley splitting in higher Landau levels is considered 
elsewhere.\cite{ando79,allmen06}  Eq.~(\ref{eq:LL}) now becomes
\begin{equation}
F_{xy}(x',y') =\frac{1}{\sqrt{2\pi l_B^2}} e^{-({x'}^2+{y'}^2)/4l_B^2} .
\end{equation}
We assume a strong electric field, so that only one quantum well interface contributes
to the valley splitting.  For a smoothly tilted interface, the valley interaction potential
is given by 
\begin{equation}
V_v(z')=v_v\delta (z') .
\end{equation}
The valley splitting is then given by
\begin{eqnarray}
E_v &\simeq & \frac{v_v}{\pi l_B^2}\left| \int \left| F_z(z') \right|^2
e^{-{r'}^2/2l_B^2}e^{2ik_0x' \vartheta} \delta(z')\, dx' dy' dz' \right| 
\nonumber \\ &=&
2v_vF_z^2(0)e^{-2(k_0l_B\theta )^2} , \label{eq:EvsB}
\end{eqnarray}
where we have used the fact that
$z=-x'\sin \vartheta +z'\cos \vartheta \simeq -x' \vartheta$, along the
$z'=0$ interface.

The preceding results are obtained for perfectly uniform steps at a quantum well 
interface, which 
we approximate by a smooth, uniform tilt.  The resulting magnetic field dependence of 
the valley splitting, first reported in Ref.~[\onlinecite{friesen06}], is shown in 
Fig.~\ref{fig:EvsB}.  The interference effect, arising from the interfacial tilt,
drives the valley splitting to zero at small
fields, as consistent with experimental observations.  However, 
the exponential suppression of $E_v$ in Eq.~(\ref{eq:EvsB}) is an 
anomalous feature caused by the absence of disorder.  If we consider more realistic
step geometries, including disorder in the step widths and profiles,\cite{swartzentruber} 
the valley splitting will be enhanced by orders of magnitude, as confirmed by 
simulations.\cite{friesen06,kharche}  

We can obtain an estimate for the valley splitting enhancement due to fluctuations
by considering a single step wiggle, as shown in the inset of Fig.~\ref{fig:EvsB}.  
The correction to Eq.~(\ref{eq:EvsB}) is 
computed by noting that the area of the left step increases by $A$, while the right
step decreases by $A$.  We have shown that the phase difference 
between neighboring steps in the valley splitting integral is $k_0a/2$.
The perturbed valley splitting integral is therefore given by
\begin{equation}
E_v\simeq |E_{v0}+4e^{i\phi}A\,v_vF_z^2(0)F_{xy}^2({\bm r}'_0)\sin (k_0a/4)| .
\label{eq:Evfluc}
\end{equation}
Here, we have expressed the unperturbed result of Eq.~(\ref{eq:EvsB}) as $E_{v0}$, and
we have assumed the amplitude of the envelope function is approximately constant  
across the wiggle at position ${\bm r}'={\bm r}'_0$.  We also note that
the two terms in Eq.~(\ref{eq:Evfluc}) enter the valley splitting integral with a phase 
difference $\phi$ that depends on ${\bm r}'_0$.
Let us approximate $F_{xy}({\bm r}'_0)\simeq 1/2\pi l_B^2$ and $\sin (k_0a/4) \simeq 1$.
Then the fluctuation contribution is 
\begin{equation}
v_vF_z^2(0)\frac{2A}{\pi l_B^2} ,
\end{equation}
which has a linear dependence on the magnetic field.  Therefore, at low fields, 
the fluctuation contribution dominates over the estimate obtained for a uniform tilt, 
Eq.~(\ref{eq:EvsB}).  More realistic
fluctuation models would include a distribution of fluctuations, with 
contributions that partially
cancel out due to interference effects.  However, the assumption of one dominant fluctuation
loop (the ``plateau" model\cite{friesen06}), leads to the
linear dependence shown in Fig.~\ref{fig:EvsB}, which is consistent with experiments.
(To obtain correspondence with Ref.~[\onlinecite{friesen06}], we note that the ``excess
area" in that paper corresponds to $2A$ in our notation.)

\section{Conclusions}
\label{sec:conclusions}
In this paper, we have developed an effective mass theory for the valley 
splitting of a strained SiGe/Si/SiGe quantum well.
To compute the valley splitting of a perfect quantum well with no steps, one needs two 
input parameters that describe the location and curvature of the band minimum as well a 
the valley coupling constant $v_v$ that captures the relevant microscopic details of the 
interfaces.  These parameters must be obtained from a more microscopic theory, or from experiments.  It is worth noting that, unlike bulk properties, the 
interface can vary from system to system, so that $v_v$ should be determined for each case.  
In this work, we have used a simple tight binding theory to compute $v_v$, as a function
of the conduction band offset $\Delta E_c$, assuming a sharp 
heterostructure interface.  The results provide
excellent agreement between the effective mass and tight binding theories.

Once $v_v$ is known, we may apply the effective mass theory to a 
range of important problems.  Here, we have considered the finite square well,
with and without an applied electric field.  We have also performed a 
self-consistent analysis of a 2DEG, using the Hartree approximation.
Excellent agreement between effective mass and atomistic theories confirms our 
main conclusion, that the 
valley splitting in this system can be fully explained 
through a single coupling constant, $v_v$.

The effective mass theory is particularly useful for two or three-dimensional
geometries, which cannot be easily treated in atomistic theories, due to
scaling constraints.  Here, we have applied the effective mass formalism to the 
inherently 2D problem of valley splitting in a magnetic field for a quantum well grown on a 
miscut substrate.  We find that interference effects strongly suppress the valley splitting 
at low magnetic fields.  However this suppression is reduced by introducing a small
amount of disorder into the step-like geometry of the quantum well interface, as consistent 
with experimental evidence and simulations.

N.B.  In the final stages of preparation of this manuscript, we became aware of
Ref.~[\onlinecite{nestoklon}], which presents some similar results to those reported
here.

\begin{acknowledgments}
We would like to acknowledge stimulating discussions with M. A. Eriksson,
G. Klimeck, A. Punnoose, 
and P. von Allmen.  This work was supported by NSA under ARO contract  
number W911NF-04-1-0389 and by the National Science Foundation through the ITR 
(DMR-0325634) and EMT (CCF-0523675) programs.
\end{acknowledgments}

\end{document}